\documentclass[unpublished]{JHEP3} 
\usepackage{latexsym}
\usepackage{epsfig}
\usepackage{amssymb}
\usepackage{amsthm}
\usepackage{amsmath}
\usepackage{amssymb,amsfonts}
\usepackage{verbatim}

\newcommand{\be}{\begin{eqnarray}}
\newcommand{\ee}{\end{eqnarray}}

\JHEPspecialurl{http://jhep.sissa.it/JOURNAL/JHEP3.tar.gz}

\usepackage{epsfig,multicol,bbm}

\newcommand\fverb{\setbox\fverbbox=\hbox\bgroup\verb}
\newcommand\fverbdo{\egroup\medskip\noindent%
            \fbox{\unhbox\fverbbox}\ }
\newcommand\fverbit{\egroup\item[\fbox{\unhbox\fverbbox}]}
\newbox\fverbbox

\title{Scalar perturbations in p-nflation: the 3-form case}

\author{Cristiano Germani\\
   LUTH, Observatoire de Paris, CNRS UMR 8102, Universit\'{e} Paris Diderot,
5 Place Jules Janssen, 92195 Meudon Cedex, France\\
    E-mail: \email{cristiano.germani@obspm.fr}}

\author{Alex Kehagias\\
   Department of Physics, National Technical University of Athens
GR-15773, Zografou, Athens, Greece\\
   E-mail: \email{kehagias@central.ntua.gr}}

\received{}       
\accepted{}       

\abstract{We calculate the primordial spectrum of scalar
perturbations of the 3-form inflation and we find that the curvature perturbations decay at late times. As as result, although
a non-minimally coupled massive 3-form field may
drive inflation at early times, it should be assisted by other fields in order to reproduce the observed
temperature fluctuations of the CMB sky.}


\begin{document}


\section{Introduction}

The latest cosmological data \cite{wmap}  agree impressively with the assumption that our Universe is, at large scales, homogeneous, isotropic and spatially flat, {\it i.e.}, that it is well described by a Friedmann-Robertson-Walker (FRW) spatially flat geometry. This observation is however a theoretical puzzle. A flat FRW Universe is in fact an extremely fine tuned solution of Einstein equations with normal matter \cite{dodelson}. In the last twenty years or so many attempts have been put forward to solve this puzzle (see for example \cite{many}-\cite{bounce}).
However, the most developed and yet simple idea still remain inflation. Inflation solves the homogeneity, isotropy and flatness problems in one go just by postulating a
rapid expansion of the early time Universe post Big Bang. Nevertheless, a fundamental realization of this idea is still eluding us. Originally,
the effective theory of inflation has been realized by sourcing General Relativity (GR) with a slow ``rolling'' massive scalar field \cite{chaotic} with minimal or even
non-minimal kinetic term \cite{k}.
Fundamental scalar fields are however not yet
discovered in nature so in principle an inflating cosmology might well be realized  by other, more complex fields.
Initiated by the idea that inflation might be driven by 1-form fields \cite{vec}-\cite{muk}
\footnote{Massive vector fields where also used to reproduce the scalar primordial perturbations 
on scalar field inflating backgrounds in \cite{dim}. Massive 3-forms minimally coupled to gravity
 have been discussed 
in \cite{kalop}.}, it has been lately showed that an inflationary scenario might be realized by using general p-forms \cite{pnflation} (see also \cite{mota} for a slightly different realization of the same idea).

In this realization of inflation, p-forms are always massive and non-minimally coupled to gravity.
 Although this last property might seems odd, non-minimal coupled fields to gravity
has been employed in various physical cases. For example, there are evidences that
a non minimal coupling of the Higgs field with gravity \cite{higgs}
might provide a very appealing particle physics scenario for inflation. Exotic fields such as higher spin fields can consistently
propagate only if non-minimally coupled
to gravity \cite{hs},\cite{hs1}. String theory compactifications
always introduce non minimal couplings  of geometric extra-dimensional fields and
four dimensional gravity \cite{hd}, \cite{hd1} and finally, interesting successful
 models of Dark Energy
are based on non minimal couplings \cite{dark}.

Although we will not here discuss whether p-nflationary models might come from a fundamental
 theory we will discuss their observational signatures in the Cosmic Microwave Background (CMB).
In particular, we will here focus on the 3-form scenario and show that, for the small field
inflationary case (for a realization as symmetry breaking scenario see \cite{natural} for natural
inflation), the spectral index of the primordial perturbations, generated by the 3-form cannot account for the latest astronomical data \cite{wmap}.

\section{Generating inflation with a 3-form}

A possible 3-form field action able to produce inflation is \cite{pnflation}
\begin{eqnarray}\label{S3}
S=\int d^4 x \sqrt{-g} \{\frac{1}{2\kappa^2} R&-&\frac{1}{48}F_{\mu\nu\rho\sigma}F^{\mu\nu\rho\sigma}-\frac{1}{12}V(
A_{\mu\nu\rho}A^{\mu\nu\rho})+\frac{1}{8}R\, A_{\mu\nu\rho}A^{\mu\nu\rho}-\cr
&-&\frac{1}{2}A_{\mu\nu\kappa}\,R^{\kappa\lambda}\,{A_\lambda}^{\mu\nu}\}
\label{action3}
\end{eqnarray}
where
\be
F_{\mu\nu\rho\sigma}=\nabla_\mu A_{\nu\rho\sigma}-\nabla_\sigma A_{\mu\nu\rho}+\nabla_\rho A_{\sigma\mu\nu}-\nabla_\nu A_{\rho\sigma\mu}\ ,
\ee
$V$ is a gauge breaking potential and $A_{\alpha\beta\gamma}$ is the 3-form field.

By varying the action (\ref{S3}) with respect to the 3-form one finds the following field equations
\be
\nabla_\kappa F^{\kappa\mu\nu\rho}-\frac{\partial V}{\partial A^{\mu\nu\rho}}+\frac{3}{2} R\, A_{\mu\nu\rho}-2(
{R^\kappa}_{\rho}
A_{\mu\nu\kappa}+{R^\kappa}_{\nu}
A_{\rho\mu\kappa}+{R^\kappa}_{\mu}
A_{\nu\rho\kappa})=0\ ,
\ee
which may be written as
\be
&&\nabla^2 A_{\mu\nu\rho}-\nabla_\sigma\nabla^\mu A_{\mu\nu\rho}+\nabla_\rho\nabla^\mu A_{\mu\nu\sigma}-
\nabla_\nu\nabla^\mu A_{\mu\rho\sigma}-\frac{\partial V}{\partial A^{\mu\nu\rho}}\nonumber \\&&+
{R_{\mu\nu}}^{\lambda\sigma}A_{\rho\lambda\sigma}+{R_{\rho\mu}}^{\lambda\sigma}A_{\nu\lambda\sigma}+
{R_{\nu\rho}}^{\lambda\sigma}A_{\mu\lambda\sigma}\nonumber\\
&&-3( {R^\kappa}_{\rho} A_{\mu\nu\kappa}+{R^\kappa}_{\nu}
A_{\rho\mu\kappa}+{R^\kappa}_{\mu} A_{\nu\rho\kappa})=0\, .
\label{a3}
\ee
The Einstein equations take the standard form
\be
R_{\mu\nu}-\frac{1}{2} g_{\mu\nu} R=\kappa^2T_{\mu\nu}\ ,
\ee
where
\be
T_{\mu\nu}&=&\frac{1}{6}F_{\mu\kappa\sigma\rho}{F_\nu}^{\kappa\sigma\rho}-\!\frac{3}{4} R
A_{\mu\kappa\sigma}{A_\nu}^{\kappa\sigma}+\frac{1}{6}\frac{\delta V}{\delta g^{\mu\nu}}
+A_{\alpha\beta\gamma}{R^\gamma}_\mu {A_{\nu}}^{\alpha\beta}+
A_{\alpha\beta\mu} R_{\gamma\nu} A^{\gamma\alpha\beta}\nonumber\\&&
+A_{\alpha\mu\gamma}R^{\gamma\beta}{{A_\beta}^\alpha}_\nu
+A_{\mu\beta\gamma}R^{\gamma \sigma}{A_{\sigma\nu}}^\beta
-\frac{1}{2}\Big{[}(\delta_\mu^\rho\delta_\nu^\gamma+\delta_\nu^\rho\delta_\mu^\gamma)
\nabla_\beta\nabla_\rho
-\delta_\mu^\gamma g_{\beta\nu}\nabla^2\nonumber\\&&
-g_{\mu\nu}\nabla_\beta\nabla_\gamma\Big{]}A_{\kappa\lambda\gamma}A^{\beta\kappa\lambda}
-\frac{1}{4} R_{\mu\nu}A_{\alpha\beta\gamma}A^{\alpha\beta\gamma}-
\frac{1}{4} g_{\mu\nu}\nabla^2 A_{\alpha\beta\gamma}A^{\alpha\beta\gamma}
+\frac{1}{4} \nabla_\mu\nabla_\nu A_{\alpha\beta\gamma}A^{\alpha\beta\gamma}\nonumber\\ &&
+g_{\mu\nu}(-\frac{1}{48}F_{\kappa\lambda\rho\sigma}F^{\kappa\lambda\rho\sigma}-\frac{1}{12}V(
A_{\kappa\lambda\rho}A^{\kappa\lambda\rho})+\frac{1}{8}R\, A_{\kappa\lambda\rho}A^{\kappa\lambda\rho}-
\frac{1}{2}A_{\kappa\lambda\sigma}\,R^{\sigma\rho}\,{A_\rho}^{\kappa\lambda})\label{tmn3}\ .
\ee
In a FRW background
\be
ds^2=-dt^2+a(t)^2 \delta_{ij}dx^i dx^j\ ,
\ee
where latin indices are for spatial directions and $\delta_{ij}$ is the spatial euclidean metric, the 3-form field is only time dependent. The components of this form may be
parameterized in terms of a 2-form $a_{ij}$ and a scalar $\phi$ as
\be A_{tij}=a_{ij}(t) \, , ~~~~~A_{ijk}=\phi(t) \epsilon_{ikj}\ ,
\label{a33} \ee
where $\epsilon_{ijk}$ is the spatial volume
element.
Then the field equations (\ref{a3}) are explicitly written
as \be
&& a_{ij}\left(\frac{\ddot{a}}{a}-\frac{\dot{a}^2}{a^2}+V''\right)=0\, , \nonumber\\
&& \ddot{\phi}+3\frac{\dot{a}}{a}\dot{\phi}+V'=0 \label{phi}\ ,
\ee
where, in this background, $V(A^2)=V(\phi)$ and $V'=dV/d\phi$.

Symmetries of the FRW background forces
\be
a_{ij}=0\ ,
\ee
which clearly solves the first of (\ref{phi}).
In this case then the degrees of freedom of the 3-form are only encoded in the scalar
$\phi$ which satisfies the equation of a potential inflation.

Plugging (\ref{a33}) in the energy-momentum tensor (\ref{tmn3}), we find that
\be
&& T_{00}=\frac{1}{2}\dot{\phi}^2+\frac{1}{2}V\, , ~~~T_{0i}=0\nonumber \\&&
T_{ij}=a^2\delta_{ij}\left(3\frac{\dot{a}}{a}\dot{\phi}\phi +\frac{1}{2}V+\frac{1}{2} \dot{\phi}^2+\phi\ddot{\phi}\right)\ .
\ee
Using (\ref{phi}) in the Einstein equations we get
\be\label{einsteinphi}
&&3 \frac{\dot{a}^2}{a^2}=\frac{1}{2}\dot{\phi}^2+\frac{1}{2}V\ ,\\
&& 2\frac{\ddot{a}}{a}+\frac{\dot{a}^2}{a^2}=-\frac{1}{2}\dot{\phi}^2+\frac{1}{2}V\ ,
\ee
which are the standard equations for a scalar field. Thus, the action for a 3-form non-minimally coupled to gravity
mimics a standard scalar field theory minimally coupled to gravity, at least at the background level. Of course, as we shall see, at the perturbative level, the 3-form inflation might greatly differs from standard inflation.

An inflationary period is, as usual, defined by the period in which slow roll conditions \cite{dodelson}:
\be
\epsilon\equiv\frac{d}{dt}\left(\frac{1}{H}\right)\ll 1\ ,\ \delta\equiv\frac{1}{H}\frac{\ddot\phi}{\dot\phi}\ll 1\ ,
\ee
where $H=\dot a/a$ and $\dot f=\partial_t f$, are satisfied.

\section{The gravitational waves and stability problems}

It has been argued in \cite{hcp} that the 1-form inflation, the so-called ''Vector-Inflation''
(VI) of \cite{muk}, might produce ghost instabilities. The same conclusion have also  been drawn
\cite{pnflation}  by
realizing that the Stuckelberg field, introduced to restore the $U(1)$ gauge symmetry of
 the massive non-minimally coupled vector field of VI to gravity, is ghost-like during
slow roll on a fixed FRW background. The same issue appears in the 2-form case, while
the 3-form, as well as the 0-form inflationary cases are exceptions \cite{pnflation}. Although a ghost-like field on a fixed gravitational background is worrisome, it is still not clear whether a consistent analysis performed by introducing the gravitational degrees of freedom into the perturbed 1 and 2-forms inflationary scenarios, might really produce sub-horizons instabilities on cosmological backgrounds \cite{vitaly}.

A second issue about the stability of p-nflation has been raised by looking at the gravitational
 wave spectrum \cite{gravwave}. In \cite{gravwave}, by assuming the decomposition theorem for
 linear perturbations, it has been argued that large field inflationary models suffer from
gravitational waves instabilities. This conclusion however, can  only be directly applied to the 3-form
 inflation, where the decomposition theorem may be used. In fact in \cite{vitaly} it has been shown that 
in the VI case, vector scalar and tensor perturbations are generically coupled. The same can
 be shown for the 2-form case. Nevertheless, such couplings are statistically suppressed for a large number ($N$) of
vector or tensor fields so that the results of \cite{gravwave} should be approximately 
(up to order $1/\sqrt{N}$ corrections)  valid\footnote{We thank S. Yokoyama and T. Kobayashi for point out 
this to us.}.

It is instructive to see why large field 3-nflationary scenarios are gravitationally unstable.
Following closely \cite{gravwave}, the perturbed metric in conformal time is
\be
ds^2=a^2(\eta)\left[-d\eta^2+\left(\delta_{ij}+\gamma_{ij}\right)dx^i dx^j\right]\ ,
\ee
where $\gamma_{ij}$ is a tensor perturbation, {\it i.e.} $\gamma^i_i=0=\partial_i \gamma^i_j$. By using the decomposition (\ref{a33}) and expanding at first order in the gravitational perturbations the action (\ref{S3}) one finds
\be
\delta S=\int \frac{a^2}{8\kappa^2}\Omega^2\left[(\gamma'_{ij})^2-c_s^2(\partial_k \gamma_{ij})^2\right]d\eta d^3x\ ,
\ee
where $'=\partial_\eta$, $\Omega^2=1+3/2 \kappa^2\phi^2$ and
\be\label{cs}
c_s^2=\frac{2-\kappa^2\phi^2}{2+3\kappa^2\phi^2}\ ,
\ee
is the sound speed of the perturbation.
From the definition (\ref{cs}) we automatically see that for large field inflationary scenarios, {\it i.e.} for $\kappa^2\phi^2\gg 1$, the gravitational perturbations become tachyonic by acquiring an imaginary sound speed. Of course in the small field case this does not happen and in fact gravitational
perturbations behaves, at lowest order in $\kappa^2\phi^2$, as in the usual scalar field (0-form) inflation.

\section{Small field 3-form inflation: primordial perturbations}

In the previous section we realized that the 3-form inflation might only be consistent in the small field formulation. For this reason we
specialize our analysis to a potential inspired by a symmetry breaking process as in natural inflation \cite{natural}
\be\label{potential}
V(A^2)=6V_0+s m^2 A^2\ , \label{va}
\ee
where $s=\pm 1$.
This potential indeed reminds an expansion of a mexican hat Higgs-like potential around the unstable point. Slow roll conditions for the potential
(\ref{potential}) are satisfied whenever $V_0\gg \frac{m^2 A^2}{6}$, also note that during slow roll $H/m \gg 1$. This last property will be crucial to
show that the produced spectrum of primordial scalar perturbations cannot be flat.

With (\ref{potential}), the action (\ref{S3}) reduces to
\be\label{S3s}
S=\int d^4 x \sqrt{-g} \{\frac{1}{2\kappa^2} R&-&\frac{1}{48}F_{\mu\nu\rho\sigma}F^{\mu\nu\rho\sigma}-s \frac{1}{12}m^2
A_{\mu\nu\rho}A^{\mu\nu\rho}+\frac{1}{8}R\, A_{\mu\nu\rho}A^{\mu\nu\rho}-\cr
&-&\frac{1}{2}A_{\mu\nu\kappa}\,R^{\kappa\lambda}\,{A_\lambda}^{\mu\nu}-2V_0\}\ .
\ee

\subsection{Dual Theory}
To study the scalar perturbations of (\ref{S3s}), it easier to work on the dual scalar formulation of the 3-form inflation \cite{pnflation}.

The action (\ref{S3s}) may be equivalently written as
\begin{eqnarray}
\!\!\!\!\!\!\!\!\!\!\!\!\! S=\int d^4 x \sqrt{-g} \left(\frac{1}{2\kappa^2} R+\frac{1}{48}F_{\mu\nu\rho\sigma}F^{\mu\nu\rho\sigma}+
\frac{1}{6}A_{\mu\nu\rho}\nabla_\sigma F^{\mu\nu\rho\sigma}-\frac{1}{2}A_{\mu\nu\kappa}
M^{\kappa\lambda}{A_\lambda}^{\mu\nu} -2V_0\right), \label{action323}
\end{eqnarray}
where
\be
M^{\kappa\lambda}= g^{\kappa\lambda}\left(s\frac{m^2}{6}-\frac{R}{4}\right)+R^{\kappa\lambda}\ .
\ee

Integrating out $F_{\mu\nu\rho\sigma}$ we get back the original action (\ref{S3s}).
The dual theory is now obtained by expressing the field strength and the 3-form potential in dual fields, {\it i.e.},
\be
F_{\mu\nu\rho\sigma}=m \epsilon_{\mu\nu\rho\sigma} \Phi\ ,
\ee
and
\be
A_{\mu\nu\rho}=\epsilon_{\mu\nu\rho\alpha}B^\alpha\ .
\ee
The dual action then reads
\be
S=\int d^4 x \sqrt{-g} \left(\frac{1}{2\kappa^2}R-\frac{1}{2}m^2\Phi^2-m B^\alpha\partial_\alpha\Phi+\frac{m^2}{2}\Delta^{\alpha\beta}
B_{\alpha} B_{\beta}\right)\ , \label{dual3}
\ee
with
\be
\Delta^{\alpha\beta}=\left(s+\frac{R}{2m^2}\right)g^{\alpha\beta}-\frac{2}{m^2}R^{\alpha\beta}\ . \label{d1}
\ee
The effective theory for the scalar field $\Phi$ is then obtained by integrating out $B_{\alpha}$. By defining
\be
\Lambda^{\alpha\mu}\Delta_{\alpha\nu}=\delta^\mu_\nu\ , \label{d2}
\ee
we have
\be\label{scalar}
S=\int d^4 x \sqrt{-g} \left(\frac{1}{2\kappa^2}R-\frac{1}{2}\Lambda^{\alpha\beta}\partial_\alpha\Phi\partial_\beta\Phi-\frac{1}{2}m^2\Phi^2-2V_0\right)\ . 
\ee
Variation of the above action with respect the metric produce the following Einstein equations
\be
R_{\mu\nu}-\frac{1}{2}g_{\mu\nu}R=\kappa^2 T_{\mu\nu}
\ee
where
\begin{eqnarray}\label{T}
T_{\mu\nu}&=& \left(1+\frac{R}{2m^2}\right)\xi_\mu\xi_\nu-\frac{\xi^2}{2m^2}R_{\mu\nu}-\frac{1}{2}g_{\mu\nu}
\left(m^2\Phi^2+\xi^\kappa\partial_\kappa \Phi\right)\nonumber \\
&&-\frac{1}{m^2}\left(\nabla^\kappa\nabla_\mu S_{\nu\kappa}+\nabla^\kappa\nabla_\nu S_{\mu\kappa}-g_{\mu\nu}
\nabla^\kappa\nabla^\lambda S_{\kappa\lambda}-\Box S_{\mu\nu}\right)\ ,
\end{eqnarray}
with
\be
\xi^\mu=\Lambda^{\mu\nu} \partial_\nu\Phi\, , ~~~~~
S^{\mu\nu}=\xi^\mu\xi^\nu -\frac{1}{4}g_{\mu\nu}\xi^2\ .
\ee
In addition, the scalar field equation may be written as
\be\label{eqmphi}
\nabla_\mu\xi^\mu=m^2 \Phi\ .
\ee
At slow roll level, {\it i.e.} by neglecting higher time derivatives of $\Phi$ into the scalar field equation (\ref{eqmphi}), we have
\be\label{Phi}
\dot \Phi\simeq-\frac{s m^2}{3H}\Phi\ .
\ee
Substituting (\ref{Phi}) into the energy momentum tensor (\ref{T}) one obtains, considering $H/m\gg 1$, in a FRW background
\be
-T^t_t+\frac{1}{3}T^i_i\simeq \frac{2 m^2}{9 H^2}\dot \Phi^2\ .
\ee
Let us now compare the same result in the variable $\phi$. By looking at the equations (\ref{einsteinphi}) we get,
\be
-T^t_t+\frac{1}{3}T^i_i\simeq 2\dot \phi^2\ ,
\ee
therefore, at slow roll level, we have
\be
\dot \Phi^2\simeq \frac{9H^2}{m^2}\dot \phi^2\ .
\ee
The same result can obviously be obtained by directly using the definition of the field strength
$F_{\alpha\beta\gamma\delta}$ in terms of $\Phi$ and $\phi$. The exact relation between $\Phi$ and $\phi$ is in fact
\be\label{relation}
m\Phi a^3=\partial_t(a^3\phi)\ .
\ee
Finally, the slow roll parameter in the $\Phi$ variable, at the lowest order in $m/H$, turns out to be
\be
\epsilon\simeq\frac{\kappa^2}{2}\frac{m^2}{9 H^4}\dot \Phi^2\ .
\ee

\subsection{The scalar spectrum}

To calculate the scalar perturbations we follow the ADM method pioneered in
 \cite{malda}. The perturbed metric in the ADM formalism might be written as
\be
ds^2=-N^2dt^2+h_{ij}\left(dx^i+N^i dt\right)\left(dx^j+N^j dt\right)\ .
\ee
Since the non minimal coupling of curvatures with the kinetic term of $\Phi$, it is wise
 here to use the $\delta \Phi=0$ gauge for scalar perturbations.
 In this gauge, scalar perturbations are completely absorbed by the gravitational potentials and can be parameterized as
\begin{eqnarray}
h_{ij}=a^2\left(1+2\zeta\right)\delta_{ij}\ ,\ N^i=\partial_i\psi=\partial_i(\psi_1+\epsilon\psi_2+\ldots)\ ,\cr N=1+\delta N=1+N_1+\epsilon N_2+\ldots\ ,
\end{eqnarray}
where the expansion in slow roll parameter has been written explicitly for 
the lapse ($N$) and the shift ($N^i$).

At zeroth order in the slow roll parameters, for small fields scenarios, we have that $\dot \Phi=0=\Phi$.
Variation of the action (\ref{scalar}) with respect to $N_1$ and $\psi$ gives \cite{malda}
\be
N_1=\frac{\dot \zeta}{H}\ ,\ \psi=-\frac{\zeta}{a^2 H}\ ,\ {\rm where}\ a=e^{Ht}\ {\rm (zeroth\ order\ in\ slow\ roll)}\ .
\ee
In principle now, since the scalar field part of the action is already at order $\epsilon$, one should consider the next to leading order expansion in the lapse and shift and vary with respect to $N_2,\psi_2$. However, only the purely gravitational action
contributes in this variation for the chosen gauge. Therefore one readily obtains \cite{malda}
\be
\delta N=\frac{\dot \zeta}{H}\ ,\ \psi=-\frac{\zeta}{a^2 H}\ ,
\ee
at first order in slow roll.

For an easy comparison with the standard perturbative analysis, we can now split the action (\ref{dual3}) into a standard
minimally coupled action of a massive scalar field $\phi$ to gravity (where $\phi=\phi(t)$),
and an additional non-minimally coupled action as follows
\be
S=S_a+S_s\ ,
\ee
where
\be\label{standard}
S_{s}=\int d^4 x \sqrt{-g} \left(\frac{1}{2\kappa^2}R-\frac{1}{2}g^{\alpha\beta}\partial_\alpha\phi\partial_\beta\phi-\frac{1}{2}m^2\phi^2-2V_0\right)\ ,
\ee
\be\label{add}
S_{a}=\int d^4 x \sqrt{-g} \left(-\frac{1}{2}\Lambda^{\alpha\beta}\partial_\alpha\Phi\partial_\beta\Phi-\frac{1}{2}m^2\Phi^2+
\frac{1}{2}g^{\alpha\beta}\partial_\alpha\phi\partial_\beta\phi+\frac{1}{2}m^2\phi^2\right)\ .
\ee
The quadratic action in linear perturbations coming from (\ref{standard}) is, at first order in $\epsilon$ \cite{malda}
\be\label{sp}
\delta S_s=\int d^4 x\frac{\dot\phi^2}{2H^2}\left[a^3\dot\zeta^2-a(\partial_i\zeta)^2\right]\ .
\ee
We turn now our attention to the additional action (\ref{add}). 

In the chosen gauge, and during inflation, the dominant part of the 
kinetic term of curvature perturbation is $s\frac{9}{2} \frac{\dot\phi^2}{m^2}a^3\dot\zeta^2$.
This shows that the case $s=-1$ generates ghost instabilities corresponding to tachyonic 
instability in the dual three-form formulation, as it can be seen in (\ref{va}). We will then 
only consider the $s=1$ case.

By using (\ref{relation}), at the lowest slow roll order,
we have\footnote{We assumed that $H/m\zeta\ll 1$.}, 
\be\label{addp}
\delta S_a=\int d^4x \frac{\dot \phi^2}{2H^2}\frac{9H^2}{m^2}
\left[a^3\dot\zeta^2-a^3 3H^2\zeta^2\right]+{\cal O}\left(\frac{3H}{m}\right)-\int d^4 x
\frac{\dot\phi^2}{2H^2}a^3\dot\zeta^2\ .
\ee
Combining now (\ref{sp}) with (\ref{addp}) we finally have
\be
\delta S\simeq\int d^4x \frac{9\dot \phi^2}{2m^2}\left[a^3\dot\zeta^2-a^3 3H^2\zeta^2-\frac{m^2}{9H^2}a(\partial_i\zeta)^2\right]\ .\label{ss}
\ee

By varying the previous action with respect to $\zeta$, we have, in conformal time $a d\eta=dt$, and by using the variable $v=a\zeta$,
\be
v''+v\left[\tilde k^2+\frac{1}{\eta^2}\right]=0\ , \label{vv}
\ee
where $\tilde k=m/3H k$ and $a=-1/(H\eta)$. In particular, it is clear from (\ref{ss}) that $m/3H$ is the sound speed of the curvature perturbations. At super horizon scales, i.e. $\tilde k\ll aH$, we then have only decaying modes
\be
\zeta\sim a^{-3/2}\ . \label{za}
\ee
This proves that the curvature perturbation decay at super-horizon scales, {\it i.e.}, 
they are diluted during the expansion of the Universe. In fact, (\ref{za}) holds for 
$k\ll \frac{3H}{m} a H$ and therefore for super-horizon scales $k\ll a H$ as well since $H\ll m$. 

The decay of curvature perturbations has a clear physical reason: In general inflationary scenarios $\zeta\approx {\rm const.}$ on large scales. This follows simply from the energy-momentum conservation as well as the adiabaticity condition. In our case however, adiabaticity is lost due to non-minimal couplings. This is the physical reason for the super-horizon decay of curvature perturbations. Moreover, it should be noted that this result is gauge-independent as it is related to the only scalar degree of freedom of the theory ($\zeta$), which is a gauge-invariant quantity.

\section{Conclusions}

The idea that p-forms coupled to gravity might be the source of inflation
 has been put forward in \cite{pnflation} generalizing vector inflation \cite{vec}-\cite{dim}
to higher order antisymmetric fields.  In particular, it has been
shown that any p-form conveniently coupled to gravity might
produce an inflationary background. In this respect, the scalar
and vector fields are only the special 0- and 1-form cases of the
general p-nflationary scenario. The mechanism of p-nflation is triggered by
 a non-minimal coupling of
the massive p-form fields to gravity making the p-form able to mimic
a slow rolling inflaton at the background level. However, as 1- and 2-nflation break isotropy, special configurations, or
 randomly distributed large number of fields, are needed for compatibility to an homogeneous and isotropic FRW background. For the standard scalar field
inflation (0-nflation) and 3-nflation, this is not
necessary, as scalars and 3-forms are compatible with isotropy. Moreover,
by studying the gravitational wave perturbations of 3-nflation in \cite{gravwave} it has been
shown that only small field 3-nflationary scenarios are gravitationally stable. Large fields
lead to imaginary speed of sound and a corresponding instability of the gravitational waves.
 We therefore studied in this paper small field 3-nflationary scenarios with potentials
inspired by symmetry breaking mechanisms, as in natural inflation \cite{natural}.

The main result of our paper is that scalar curvature perturbations in 3-form
inflation are decaying at super-horizon scales. This implies that the 3-form perturbations cannot be responsible for the temperature
fluctuations of the CMB sky. This is due to the fact that,
the non-minimal couplings of curvatures with the dual scalar of the 3-form inflaton,
are non adiabatic components of the linear perturbations. To reproduce the latest astronomical data then,
the 3-form inflation should be assisted by other fields, subdominant at the background level,
able to produce a growing, almost scale invariant spectrum, of perturbations. This important task if left for future work.

\acknowledgments

CG is in debt with Misao Sasaki for finding out an important mistake in the first version of this manuscript and for decisive discussions.
CG wishes to thank David Langlois and Viatcheslav Mukhanov for enlightening discussions on cosmological perturbations and gauge invariance
related to p-nflation. CG also wishes to thank Antonio Riotto for comments on the adiabatic properties of p-nflation. Finally CG wishes to thank the LMU
in Munich for hospitality during the preparation of the second version of this paper.
AK wishes to thank support from the PEVE-NTUA-2007/10079 program. This work is partially supported by the European
Research and Training Network MRTPN-CT-2006 035863-1.

\appendix

\section{Appendix}

Here we show that the number of scalar degrees of freedom for the coupled dual 3-form inflation and gravity at linear level is only one, {\it i.e.}
$\zeta$, as in standard GR.

In the gauge $\delta\Phi=0$, the only non-minimal term which might contain a dynamics for $\delta N$ or $\psi$ is
\be
L\propto \sqrt{-g}\delta R\frac{\dot\Phi^2}{\bar N^2}\ ,
\ee
where $\bar N$ is the background lapse. We can study this term in generality by considering the non-linear action
\be
A=\int d^4x\sqrt{-g} R f(t)\ ,
\ee
where $f(t)$ is an external function of time which will be lately taken as $\dot \Phi^2/\bar N^2$. This action looks very much as the Einstein-Hilbert
(EH) action, however, since the presence of the function $f(t)$, the boundary terms of the (EH) action are no longer boundaries here. In ADM variables
we in fact have the residual action (after integrating out a boundary term)
\be
A_r=\int d^4x \sqrt{-g}\left[\nabla_\beta n^\beta n^\alpha\nabla_\alpha f(t)-n^\alpha\nabla_\alpha n^\beta\nabla_\beta f(t)\right]\ ,
\ee
where $n_\alpha=-N\delta^t_\alpha$.

Explicit calculation of the above action gives
\be
A_r\sim \int d^4x\sqrt{h} N\!\!&&\!\!\!\left(\left[\frac{1}{N^2}\Big{(}
2 \nabla_i N_j-\partial_t h_{ij}\Big{)}h^{ij}
-\frac{1}{2N^4}\Big{(}N^i\partial_i(N_k N^k-N^2)-2 N^i N^j \nabla_iN_j\Big{)}\right]
\dot{f}
\right.\nonumber \\
&&\left.+\frac{1}{2N^2}
\ddot{f}
\right)\ .
\ee

We see that $N,N^i$ are not dynamical
indicating that no extra dynamical degrees of freedom appear at higher orders in the slow roll parameters.

\end{document}